\documentstyle[11pt,moriond,epsfig]{article}

\def\bull{\parskip=0pt\par\noindent\hangindent=3pc\hangafter=1 $\bullet$~}

\def\kms{\ifmmode{\,\hbox{km}\,s^{-1}}\else {\rm\,km\,s$^{-1}$}\fi}

\def\hmpc{\ifmmode{h^{-1}\,\hbox{Mpc}}\else{$h^{-1}$\thinspace Mpc}\fi}

\def\et{{\it et~al.}~}

\def\sigp{\ifmmode{\sigma_p}\else {$\sigma_p$}\fi}
\def\sig1{\ifmmode{\sigma_1}\else {$\sigma_1$}\fi}
\def\r200{\ifmmode{r_{200}}\else {$r_{200}$}\fi}
\def\spose#1{\hbox to 0pt{#1\hss}}
\def\lta{\mathrel{\spose{\lower 3pt\hbox{$\mathchar"218$}}
     \raise 2.0pt\hbox{$\mathchar"13C$}}}
\def\gta{\mathrel{\spose{\lower 3pt\hbox{$\mathchar"218$}}
     \raise 2.0pt\hbox{$\mathchar"13E$}}}

\def\apj{ApJ}
\def\apjl{ApJ(Lett)}

\def\apjs{ApJS}

\input{epsf}
\input{rotate}

\def\altaffilmark#1{$^{#1}$}

\begin{document}
\vspace*{4cm}
\title{The $\Omega_M-\Omega_\Lambda$ Constraint from CNOC Clusters}

\author{
R.~G.~Carlberg\altaffilmark{1,2,8},
H.~K.~C.~Yee\altaffilmark{1,2},
E.~Ellingson\altaffilmark{1,3},
S.~L.~Morris\altaffilmark{1,4},
H.~Lin\altaffilmark{1,2},
M.~Sawicki\altaffilmark{1,2},
D.~Patton\altaffilmark{1,6},
G.~Wirth\altaffilmark{1,6},
R.~Abraham\altaffilmark{1,4,5},
P.~Gravel\altaffilmark{1,2},
C.~J.~Pritchet\altaffilmark{1,6},
T.~Smecker-Hane\altaffilmark{1,4,7}
D.~Schade\altaffilmark{4},
F.~D.~A.~Hartwick\altaffilmark{6}
J.~E.~Hesser\altaffilmark{4},
J.~B.~Hutchings\altaffilmark{4},
\& J.~B.~Oke\altaffilmark{4}
}

\address{
$^{1}$Visiting Astronomer, Canada--France--Hawaii Telescope, 
        which is operated by the National Research Council of Canada,
        le Centre National de Recherche Scientifique, and the University of
        Hawaii.
$^{2}$Department of Astronomy, University of Toronto, 
        Toronto ON, M5S~3H8 Canada.
$^{3}$Center for Astrophysics \& Space Astronomy,
        University of Colorado, CO 80309, USA.
$^{4}$Dominion Astrophysical Observatory, 
        Herzberg Institute of Astrophysics,     
        National Research Council of Canada,
        5071 West Saanich Road,
        Victoria, BC, V8X~4M6, Canada.
$^{5}$Institute of Astronomy, 
        Madingley Road, Cambridge CB3~OHA, UK.
$^{6}$Department of Physics \& Astronomy,
        University of Victoria,
        Victoria, BC, V8W~3P6, Canada.
$^{7}$Department of Physics \& Astronomy,
        University of California, Irvine,
        CA 92717, USA.
$^{8}${\it on leave at:} Carnegie Observatories, 813 Santa Barbara Street,
Pasadena, CA 91101, USA.
}

\maketitle

\abstracts{ The CNOC redshift survey of galaxy clusters measures
$\Omega_M$ from $\Omega_e(z)= M/L \times j/\rho_c$ which can be
applied on a cluster-by-cluster basis. The mass-to-light ratios,
$M/L$, are estimated from rich galaxy clusters, corrected to the field
population over the $0.18\le z \le 0.55$ range. Since the luminosity
density depends on cosmological volumes, the resulting $\Omega_e(z)$
has a strong dependence on cosmology which allows us to place the
results in the $\Omega_M-\Omega_\Lambda$ plane. The resulting
$\Omega_M$ declines if $\Omega_\Lambda>0$ and we find that
$\Omega_\Lambda<1.5$.}

\section{Introduction}

In the Friedmann-Robertson-Walker solution for the structure of the
universe the geometry and future of the expansion uniquely depend on
the mean mass density, $\rho_0$, and a possible cosmological constant.
It is a statement of arithmetic\,\cite{oort} that at redshift zero
$\Omega_M \equiv\rho_0/\rho_c= M/L \times j/\rho_c$, where $M/L$ is
the average mass-to-light ratio of the universe and $\rho_c/j$ is the
closure mass-to-light ratio, with $j$ being the luminosity density of
the universe.  Estimates of the value of $\Omega_M$ have a long
history with a substantial range of cited results\,\cite{bld}. Both
the ``Dicke coincidence'' and inflationary cosmology would suggest
that $\Omega_M=1$.  The main thrust of our survey is to clearly
discriminate between $\Omega_M=1$ and the classical, possibly biased,
indicators that $\Omega_M\simeq 0.2$.

Rich galaxy clusters
are the largest collapsed regions in the universe and are ideal to
make an estimate of the cluster $M/L$ which can be corrected to the
value which should apply to the field as a whole.  To use clusters to
estimate self-consistently the global $\Omega_M$ we must, as a
minimum, perform four operations.
\bull{Measure the total gravitational mass within some radius.}
\bull{Sum the luminosities of the visible galaxies within the same
        radius.}
\bull{Measure the field luminosity density at the cluster redshift.}
\bull{Understand the 
        differential luminosity and density evolution between
        the clusters and the field.}

\section{Survey Design}

The Canadian Network for Observational Cosmology (CNOC) designed
observations to make a conclusive measurement of $\Omega_M$ using
clusters\,\cite{yec}.  The clusters are selected from the X-ray
surveys, primarily the Einstein Medium Sensitivity
Survey\,\cite{emss1,emss2,gl}, which has a well defined flux-volume
relation. The spectroscopic sample, roughly one in two on the average,
is drawn from a photometric sample which goes nearly 2 magnitudes
deeper, thereby allowing an accurate measurement of the selection
function.  The sample contains 16 clusters spread from redshift 0.18
to 0.55, meaning that evolutionary effects are readily visible, and
any mistakes in differential corrections should be more readily
detectable.  For each cluster, galaxies are sampled all the way from
cluster cores to the distant field. This allows testing the accuracy
of the virial mass estimator and the understanding of the differential
evolution process.  We introduce some improvements to the classical
estimates of the velocity dispersion and virial radius estimators,
which have somewhat better statistical properties.  A critical element
is to assess the errors in these measurements.  The random errors are
relatively straightforward and are evaluated using either the
statistical jackknife or bootstrap methods\,\cite{et}, which follow
the entire complex chain of analysis from catalogue to result.  The
data are designed to correct from the $M/L$ values of clusters to the
field $M/L$.

\section{Results}

We find\,\cite{profile} that $\Omega_M=0.19\pm0.06$ (in a
$\Omega_\Lambda=0$ cosmology, which is the formal $1\sigma$ error.  In
deriving this result we apply a variety of corrections and tests of
the assumptions.  \bull{The clusters have statistically identical
$M/L$ values, once corrected for evolution\,\cite{global}.}
\bull{High luminosity cluster and field galaxies are evolving at a
comparable rate with redshift, approximately one magnitude per unit
redshift.}  \bull{Cluster galaxies have no excess star formation with
respect to the field\,\cite{a2390,balogh}. } \bull{Cluster galaxies
are 0.1 and 0.3 magnitudes fainter than similar field
galaxies\,\cite{profile,schade_e,schade_d,lin}.}  \bull{The virial
mass overestimates the true mass of a cluster by about 15\%, which can
be attributed to the neglect of the surface term in the virial
equation\,\cite{profile}.}  \bull{There is no significant change of
$M/L$ with radius within the cluster\,\cite{profile}.}  \bull{The mass
field of the clusters is remarkably well described by the
NFW\,\cite{nfw} profile, both in shape and scale radius\,\cite{ave}.}
\bull{The evolution of the number of clusters per unit volume is very
slow, in accord with the PS\,\cite{ps} predictions for a low density
universe\,\cite{s8}.}  \bull{The clusters have statistically identical
efficiencies of converting gas into stars, which is consistent
with the
value in the field\,\cite{omb}.}  \par\noindent These results rule out
$\Omega_M=1$ in any component with a velocity dispersion less than
about 1000~\kms.

\section{$\Omega_\Lambda$ dependence}

The luminosity density, $j$, contains the cosmological volume element
which has a very strong cosmology dependence.  The cosmological
dependence of the $\Omega_e(z)$ can be illustrated by expanding the
cosmological terms to first order in the redshift, $z$,
\begin{equation}
\Omega_e(z) \simeq \Omega_M [1 + {3\over 4}(\Omega_M^i
        -\Omega_M+2\Omega_\Lambda)z],
\label{eq:ao}
\end{equation}
where $\Omega_M$ and $\Omega_\Lambda$ are the true values and
$\Omega_M^i$ with $\Lambda=0$ is the cosmological model assumed for
the sake of the calculation\,\cite{lambda}. If there is a non-zero
$\Lambda$ then $\Omega_e(z)$ will vary with redshift. The available
data are the CNOC1 cluster $M/L$ values and the 3000 galaxies of the
preliminary CNOC2\,\cite{cnoc2_pre} field sample for $j$.  To provide
a well defined $\Omega_e(z)$ we limit both the field and cluster
galaxy luminosities at $M_r^{k,e}\le -19.5$ mag, which provides a
volume limited sample over the entire redshift range.  A crucial
advantage is of using high luminosity galaxies alone is that they are
known to have a low average star formation rate and evolve slowly with
redshift, hence their differential corrections are small, and
reasonably well determined\,\cite{profile}.  The results are displayed
in Figure~1. The fairly narrow redshift range available does not
provide a very good limit on $\Omega_\Lambda$, although values
$\Omega_\Lambda>1.5$ are ruled out. The power of this error ellipse is
to use it in conjunction with other data, such as the SNIa results
which provide complementary constraints on the
$\Omega_M-\Omega_\Lambda$ pair.

\bigskip
\epsfysize 8truecm
\centerline{\epsfbox{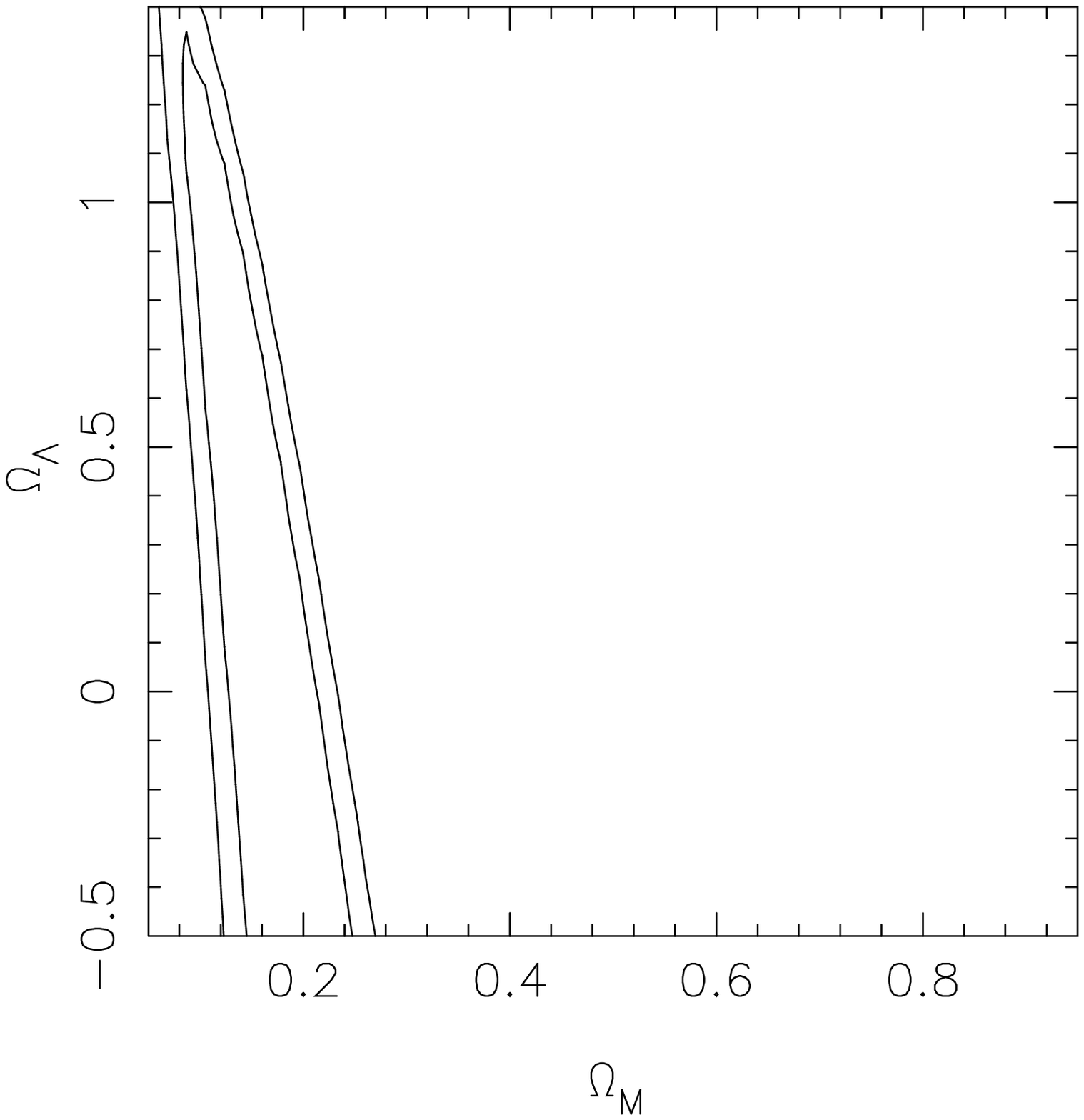}}
\noindent
{\footnotesize Figure 1: The CNOC1 cluster $M/L$ values combined with
the CNOC2 measurements of $j$ for $M_r^{k,e}\le -19.5$ mag galaxies,
gives an $\Omega_e(z)$ which leads to the plotted $\chi^2$ (68\% and
90\% confidence) contours.}
\medskip

The limit on the $\Omega_M-\Omega_\Lambda$ pair in Figure~1 has been
corrected for known systematic errors which are redshift independent
scale errors in luminosity and mass. The high luminosity galaxies in
both the cluster and field populations are evolving at a statistically
identical rate with redshift, which is close passive evolution.  If
the cluster galaxies are becoming more like the field with redshift
({\it i.~e.} the Butcher-Oemler effect, which is partially shared with
the field), so that they need less brightening to be corrected to the
field, then that would raise the estimated $\Omega_\Lambda$, although
the correction is so small that the correction would be
$\Delta\Omega_\Lambda\simeq0.3$ over this redshift interval.  The
results are completely insensitive to galaxy merging that produces no
new stars. The data indicate that there is no excess star formation in
cluster galaxies over the observed redshift range, with galaxies
fading as they join the cluster\,\cite{profile,balogh}.  The fact that
evolution of the high luminosity field galaxies is very slow and
consistent with pure luminosity evolution\,\cite{profile} (Lin, \et\
in preparation) gives us confidence that the results are reasonably
well understood. It will be very useful to have data that
extends to both higher and lower redshift, which would allow a
measurement of $\Omega_\Lambda$ and better constraints on any
potential systematic errors.

\end{document}